\documentclass[iop]{emulateapj}
\usepackage{graphicx}
\usepackage{pstricks}
\usepackage{txfonts}
\usepackage{amssymb}
\usepackage{natbib}
\usepackage{epstopdf}

\begin{document}
\title{On the Differential Rotation of Massive Main-Sequence Stars}
\author{T.M. Rogers}
\affil{Department of Mathematics and Statistics, Newcastle University,
  UK}
\affil{Planetary Science Institute, Tucson, AZ 85721, USA}

\begin{abstract}
To date, asteroseismology has provided core to surface differential
rotation measurements in eight main-sequence stars.  These stars, ranging in mass from
$\sim$1.5-9$M_\odot$, show rotation profiles ranging from uniform to
counter-rotation.  Although they have a variety of masses, these stars all have
convective cores and overlying radiative regions, conducive to
angular momentum transport by internal gravity waves (IGW).  Using
two-dimensional (2D) numerical simulations we show that angular momentum
transport by IGW can explain all of these rotation profiles.  We
further predict that should high mass, faster rotating stars be observed, the
core to envelope differential rotation will be positive, but less than
one.  
\end{abstract}


\section{Introduction}
Rotation is a key property of stars that has important consequences
for their long term evolution and eventual demise.  Rotation is
particularly important in massive stars where it contributes
significantly  to chemical mixing \citep{zahn92,talon97} and may determine the eventual explosion
energy and nucleosynthetic yield of the star \citep{heger00}. Given
its importance, it would be extremely beneficial if constraints could
be placed on stellar internal rotation as well as on the dominant
physical mechanisms responsible for such rotation.  However, theoretically
determining the internal rotation of stars is plagued by complicated
hydrodynamic processes which are difficult to simulate numerically and until recently, observations have provided
little constraint.  

Fortunately, the observational landscape has recently changed due to space
missions like Convection, Rotation
and planetary Transits (CoRoT) and Kepler.  With the continuous duty
cycle provided by these missions observers have been able to
place constraints on the internal rotation of hundreds of evolved stars using mixed modes
\citep{beck12,mosser12,deheuvels15}.  The overall 
consensus of these observations is that angular
momentum coupling between the contracting core and envelope is far more efficient
than previously expected.  Although it is still unclear what the
physical mechanism is that causes this efficient coupling, internal
gravity waves (IGW) are a key contender \citep{fuller14}.

Core-envelope differential rotation has also been measured in eight
intermediate and massive main sequence stars using both p- and
g-modes
\citep{aerts03,pam04,briquet07,kurtz15,saio15,triana15,valentina15}.
We note that the term core here is used to mean the
  region just outside the convective core and is really the inner
  radiative region.  Throughout this text we will use this
  terminology for consistencies sake but emphasize that ``core'' used here
  does \textit{not} refer to the \textit{convective} core. This handful of observations
have shown a variety of differential rotation profiles.  The
measurements of HD 157056 \citep{briquet07}, KIC 9244992
\citep{kurtz15}, KIC 11145123 \citep{saio15} and the binary system
KIC10080943 \citep{valentina15} show fairly uniform
rotation, although notably not \textit{exactly} uniform.  On
the other hand, HD29248 \citep{pam04} and HD129929 \citep{aerts03}
show cores spinning more rapidly than their envelopes.  Finally, 
HD 10526294 \citep{triana15} shows an envelope spinning faster than
the core and with the opposite direction.

It has been shown previously that IGW generated by convection are very
efficient at transporting angular momentum \citep{rg13}.  This is particularly true
in stars which have convective cores and extended overlying radiative regions.  In this configuration IGW generated at the
convective-radiative interface propagate outward into a region whose
density is decreasing dramatically. This causes wave amplitudes to
increase rapidly. Because of
this increase in amplitude, very small amplitude perturbations at
generation can lead to large perturbations in the envelope and
therefore, lead to efficient angular momentum transport if the waves
dissipate \citep{rg13}. 

Convection generates both prograde and retrograde waves at the
convective-radiative interface.  The initial symmetry breaking of a
uniformly rotating medium caused by the dissipation of predominantly
prograde or retrograde waves at the surface is a stochastic process,
meaning the angular momentum transport by IGW could either speed up
(if prograde waves are dissipated) or slow down (if retrograde waves
are dissipated) the radiative region.  

This initial symmetry breaking sets the stage for further angular velocity evolution, which will
depend on the dominant dissipation mechanism.  If waves dissipate
through nonlinear wave breaking, then subsequent evolution can vary in
sign and a strong mean flow may not develop.  If waves dissipate
predominantly through radiative dissipation then whichever sign
flow dominates initially will grow and eventually reverse in time (as
in the Quasi-Biennial Oscillation \citep{baldwin01}, although it is
still unclear whether such an oscillation would proceed in a massive
star \citep{rg13}).  Similarly, if a
critical layer develops, then any initial mean flow will be amplified,
but on a much faster timescale (than radiative diffusion alone). In
massive stars the density stratification is such that
waves are likely to non-linearly break. However, whether or not a
critical layer develops will depend on the surface wave flux, which depends on
the convective flux, and the details of the stratification.  Therefore, the outcome of IGW
transport can vary from simple efficient angular momentum transport
between the convective and radiative regions, to strong
differential rotation if a critical layer develops.  

While the observed stars vary in mass they share the common characteristic of
having convective cores with overlying radiative regions, albeit
with different extent.  Given 
the limited number and resolution of the observations, here we use a single fiducial model of a star with a
convective core and radiative envelope.  By simply varying the initial
rotation rate (to mimic different initial conditions) and
convective flux (to mimic different masses and ages) we
show that angular momentum transport by convectively driven IGW can explain
the variety of observed rotation profiles.    

\section{Observations of Core-Envelope Differential Rotation in
  Massive Main Sequence Stars}

To date core-envelope differential rotation has been measured in eight
main sequence intermediate and massive
stars.  Here we briefly summarize those results.  The
first measurement of core-envelope differential rotation ( $\Omega_{c}/\Omega_{e}$) in a main
sequence star was done by \cite{aerts03} for the B3V star
HD129929.  That star was found to have a core rotating approximately
3.6 times faster than its envelope \citep{dupret04,aerts08}.  HD29248, another
B star of similar mass \citep{pam04,aus04}, was found to have a core
spinning approximately 5 times faster than its envelope.
\cite{briquet07} found a rotation profile consistent with uniform
rotation for the $\sim$8$M_\odot$ star HD157056.  More recently,
\cite{kurtz15} and \cite{saio15} have found nearly uniform rotation
for the F stars KIC9244992 and KIC11145123.  Though importantly,
with high confidence, they find that KIC9244992 has an envelope
rotating slightly slower than its core ($\Omega_{c}/\Omega_{e} =0.97$)
and conversely KIC11145123 has an envelope spinning slightly faster
than its core ($\Omega_{c}/\Omega_{e}=1.03$).   Similarly,
\cite{valentina15} constrained core-envelope differential rotation in
both components of the binary system KIC10080943 and found one object
has a slightly faster core than envelope, while the other member shows the
opposite.\footnote{It is very likely that \textit{tidally induced}
  waves, which we do not consider here, could contribute to this rotation profile.} Finally, \cite{triana15}
 used 19 g-mode multiplets to do a full inversion to find the radial differential rotation profile
for KIC10526294.  They found that the envelope was spinning
significantly faster than the core
($\Omega_{c}/\Omega_{e} =0.3$) but perhaps more surprising, with the
opposite sign.  That is, the envelope is rotating in the opposite
direction to the core, see Fig.\ref{sant}.  

With the exception of \cite{triana15}, all of these stars only have a
measurement of the ratio of $\Omega_{c}/\Omega_{e}$ and not an
actual rotation profile.  These ratios are derived from multiplets of g-modes, which are confined to the
region just outside the convection zone, and multiplets of
p-modes, which are confined to the surface regions.  Therefore, the
differential rotation measurement is really a measure of two
regions of the star, just outside the convective core and just
beneath the surface.  Furthermore, mode identification is much easier
in slower rotators, therefore, all of the observed stars are slow rotators,
perhaps unusually so.  Therefore, these observed stars may not represent the rotation
profiles of intermediate and massive main sequence stars as a whole.  Consequently, in the following
we will consider a variety of initial rotation rates.

\section{Modeling angular momentum transport by IGW}
In order to model angular momentum transport by IGW in stellar
interiors we solve the Navier-Stokes equations in the anelastic
approximation \citep{gough69,rg05}.  The equations are solved in two-dimensions (2D),
representing an equatorial slice of the star.  Here we use a
3$M_{\odot}$ star as a fiducial model.  The radial domain extends from
0.01$R_{\star}$ to 0.90$R_{\star}$, encompassing both the convective
core and radiative envelope to accurately model the convective
generation of waves.  The reference state thermodynamic variables are
calculated from a polynomial fit to a
one-dimensional model calculated using the Cambridge stellar evolution code STARS
for a 3$M_{\odot}$ star \citep{egg71}, with a central hydrogen
fraction, X$_{c}$ =0.47.  At this age the convective core occupies
0.30$R_{\odot}$ or 14\% of the radial domain. Because of the steep
density gradient 90\% of the angular momentum of the star resides
within 60\% of the radius.
\begin{figure}
\centering
\epsscale{1.00}
\plotone{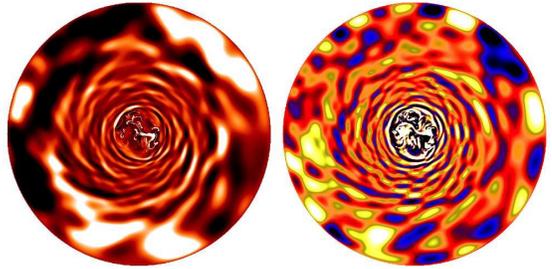}

\caption{Time snapshot of M4. (Left) Temperature perturbation with white
  hot and black cool perturbations.  (Right) Vorticity with black
  negative vorticity and white positive.}
\label{snap}
\end{figure}

These simulations, like all hydrodynamic simulations, require higher than realistic diffusion
  coefficients for numerical stability (here we use $\nu=4\times 10^{13}$ and
$\kappa=5\times 10^{11} cm^{2}s^{-1}$).  Such diffusion coefficients
  would damp IGW unrealistically on their journey to the surface of
  the star.  To compensate for this enhanced diffusion we force the
  waves harder by forcing the convection harder so
that the waves reach the surface with more realistic amplitudes.
Forcing the convection harder leads to convective velocities which are
$\sim
10-20$ times larger than expected from mixing-length theory, depending
on the model.  However, if we compare our simulated surface velocities
to those calculated assuming mixing length theory for the convective
velocities, proper density stratification and realistic diffusivities,
those velocities are comparable. More details of the numerical model
can be found in \cite{rg13}. 

Because the observations span a range of masses, here we consider slightly
different convective fluxes as a crude way to mimic different stellar
masses and ages.  Broadly, we expect higher convective fluxes
(Q/c$_v$) to be
associated with more massive stars, but given the other details we
have neglected, such as varying stratification and age, this is not
necessarily the case.  We further consider different rotation rates to
mimic different initial conditions.  Table 1
lists the initial conditions and parameters of the models considered
along with their resulting core-envelope differential rotation.
Fig.\,\ref{snap} shows a typical
 time snapshot within the
simulated domain of the temperature and vorticity for model M4. 

\begin{table}
   \centering
  \begin{tabular}{lllll}
      \hline
      Model & $\Omega_{i}$ (rad/s) & $\overline{Q}/c_{v}$
    & $\Omega_{c}/\Omega_{e}$&$\langle AM \rangle$/$AM_{i}$  \\
      
      \hline
    M1  & $10^{-7}$ & 1.5 & 2.5$\pm$2.3&1.005\\
    M2  & $5\times10^{-7}$ & 1.5 & 0.5$\pm$1.9 &1.0210\\
    M3  & $10^{-6}$  &1.5 & 3.73$\pm$3.46 & 1.010\\
    M4  & $10^{-6}$ & 3.0 & -0.06$\pm$ 0.14&1.021\\
    M5  &$ 5 \times 10^{-6}$ & 1.5 & 0.60$\pm$0.81 &0.998\\
    M6  & $5 \times 10^{-6}$ & 2.2 &1.24$\pm$1.56&1.020\\
    M7  & $5 \times 10^{-6}$ & 3.0 & -0.14$\pm$0.65&0.990\\
    M8 & $10^{-5}$&1.5 & 1.06$\pm$0.34&1.000\\
    M9 & $10^{-5}$& 3.0 & 0.20$\pm$0.11&1.010\\
    M10 & $4\times 10^{-5}$&1.5& 0.97$\pm$0.10&1.001\\
    M11 & $4\times 10^{-5}$&3.0& 0.21$\pm$0.18&1.008\\
    M12 & $8\times 10^{-5}$&1.5& 0.93$\pm$0.06&1.000\\
    M13 & $8\times 10^{-5}$&3.0& 0.12$\pm$0.03&1.011\\
    
           \hline
    \end{tabular}
   \normalsize
    \caption{Model parameters. $\Omega_{i}$ is the initial rotation
      rate given in
      rad/s. $\overline{Q}/c_{v}$ represents the
      convective forcing in units K s$^{-1}$, where $c_{v}$ is the
        specific heat at constant volume.  The values 1.5 and
        3 result in root mean squared convective velocities of $\sim$ 2.9 and 4.5
        km s$^{-1}$, respectively, values $\sim$10 and $\sim$20 times larger than predicted by mixing length theory.  The differential rotation, $\Omega_{c}/\Omega_{e}$, represents
      the mean ratio of core to envelope rotation. The time and spatial averaging
      are discussed in the text. Errors quoted are due to variations in time, also
      discussed in the text. $\langle AM \rangle$/$AM_{i}$ represents
      the integrated angular momentum compared to the initial angular
      momentum content of the system, demonstrating the level at which
    angular momentum is conserved in the system.}

\end{table}

\section{Simulated Core-Envelope Differential Rotation}

\begin{figure}
\centering
\includegraphics[width=3.6in]{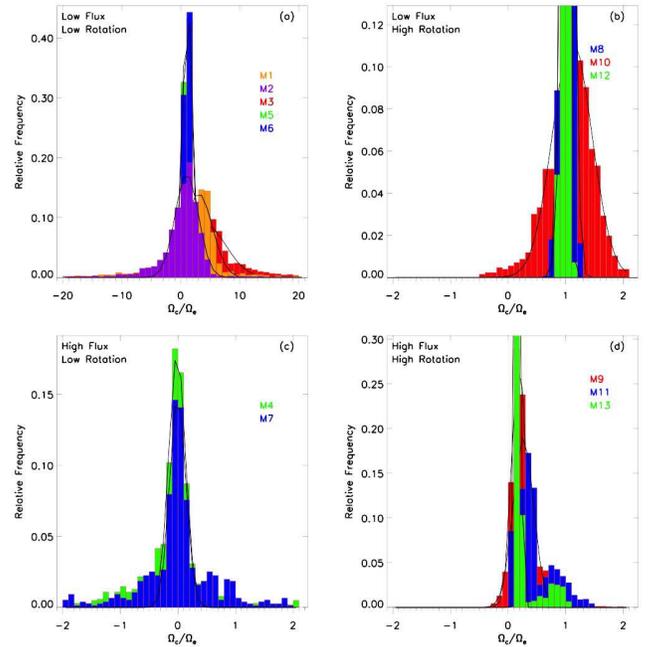}

\caption{Histogram of $\Omega_c/\Omega_e$ for models listed in Table
  1.  (a) Low flux, low rotation models M1,M2,M3,M5,M6.  Mean values
  range from $\sim$0.5 to $\sim$4. In general, these models could
  explain the rotation measurements of HD129929, HD29248, HD157056, KIC9244992, KIC11145123 and KIC10080943. (b) Low flux, high rotation models
  M8,M10,M12.  Such stars have not yet been observed.  Ratios are $\sim$1, IGW transport some angular
  momentum but not so much to bring about substantial differential
  rotation. (c) High flux, low rotation models M5 and M8.  In these models IGW transport
  significant angular momentum, causing the envelope to spin substantially faster
  than the core.  In these low rotation models, negative ratios
  (envelope spinning retrograde) are favored, nicely explaining the
  observation of \cite{triana15}. (d) High flux, high rotation models
  M9,M11 and M13.  IGW transport is efficient enough to cause
  the envelope to spin faster than the core, so that
  $\Omega_c/\Omega_e <1$, but in contrast to the slowly rotating
  models, these favor positive (prograde) surface rotation.  Again
  these stars have yet to be observed.}
\label{histogram}
\end{figure}

To mimic the regions probed by observations we average the 
core rotation over $\sim 0.2R_{\star}$ outside the convection zone
($\Omega_{c}$) and the surface rotation over $\sim 0.1 R_{\star}$
below the surface ($\Omega_{e}$).  The ratio of these two values
varies substantially in time, therefore, to best illustrate the results we show the histogram of values obtained for
each model in Fig.\ref{histogram} (details are in the figure
caption).  The distribution of values seen in Fig.\ref{histogram}
indicates the stochastic nature of wave generation and dissipation.
Hence, while most of the profiles are Gaussian with a clear average,
there is some deviation and skewness.  The values of
$\Omega_{c}/\Omega_{e}$ quoted in Table 1 are the mean values with
errors of one standard deviation.  The variability due to differences
in spatial averaging are smaller than those in time, so long
as $\Omega_{c}$ is measured within the radiative region and away from
convective overshoot.  If the core value includes the convection
zone, the ratio $\Omega_c/\Omega_e$ becomes significantly more
variable, tends to increase and its distribution is often
not Gaussian.  This may be due to inadequate time resolution or
reduced dimensionality, but is more likely due to the stochastic
nature of turbulent convection.  Each of the models is run
for at least 20 wave crossing times of the entire radiative envelope
for a typical wave (horizontal wavenumber 10 and frequency 10$\mu$Hz),
or $\sim$ 100 convective turnover times, which amounts
to $\sim 10^{7}s$.  We note that some models are run substantially
longer and do not show substantial variation and certainly none
outside the error bars quoted.

In Fig.\ref{histogram} we immediately see that the range of
differential rotation profiles seen in the simulations ($-0.03$--5) is similar to
that observed ($-0.3$--5).  More specifically, our low flux models with a variety of low rotation
rates converge to core-envelope differential rotation
values between $\sim$ 1--5, similar to seven of the eight observations of
differential rotation (HD129929, HD29248, HD157056, KIC9244992,
KIC11145123, KIC10080943). These models show
a slight preference toward values closer to one than to five,
similar to the observations. Simply, we expect
  HD129929, HD29248 and HD157056 to be described by high flux rather
  than low flux models.  However, numerous other effects (such as
  stratification, Brunt-Vaisala barrier, etc. - see
  Discussion) could affect the surface flux of waves contributing to
  these stars appearing more like low flux models.  Low
flux, high rotation models also show values very close to one.
Notably though, the averages are not exactly one.  Therefore, in these
low flux models, there is
some angular momentum transport by waves but not enough to bring the
system significantly away from its initially uniform state.  This is
particularly true in faster rotating models, where wave transport is
less efficient.   

High flux models on the other hand converge to core-envelope
differential rotation values generally between $\pm$1. In
this case, IGW are particularly efficient at spinning up (in
amplitude) the radiative envelopes and hence, envelopes generally spin
faster than cores.  For slow rotators this transport is efficient
enough and predominantly due to retrograde waves, so that negative
values are common.  On the other hand, in fast rotators, prograde waves
dominate and bring about a fast, but positive rotation in the
envelope.  It is worth noting that, in general, slow rotators tend to
favor retrograde wave deposition at the surface and hence, retrograde
envelope rotation,  while fast rotators favor prograde wave deposition
at the surface and hence, prograde envelope rotation.  At the moment, the theoretical reason for
this tendency is unknown.    

\begin{figure}
\centering
\includegraphics[width=3in]{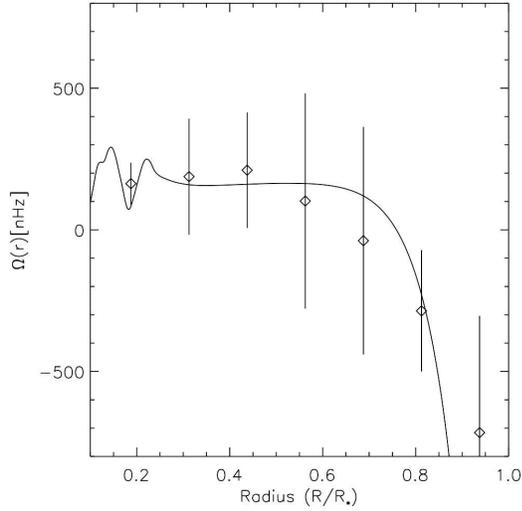}
\caption{Angular velocity as a function of radius for KIC10526294 as
  measured by \cite{triana15} (diamonds connected by solid line) with
vertical lines indicating error bars.  Time average of the angular
velocity from Model M4 is shown overlaid.}
\label{sant}
\end{figure}

The high flux, slow rotator behavior seen in these simulations is similar to the 
differential rotation pattern observed in the star 
KIC10526294 \citep{triana15}.  In Fig.\ref{sant} we show the rotation profile inferred for
KIC10526294 \cite{triana15}, with error bars, along with the time-averaged
rotation profile from M4 which was initiated with a rotation
  rate similar to KIC10526294 and which develops a counter-rotating envelope.  There we see
that a low rotation model with high IGW flux could reproduce the
observed rotation profile of KIC10526294.  The outer layers of
  the star, that are spinning retrograde, represent only $\sim$1\% of
  the angular momentum.  Therefore, those models which
  only conserve angular momentum to $\sim$2\% might not accurately
  capture the surface dynamics.  However, in this particular case (M4), the
  angular momentum is \textit{larger} than the original value, so the
  angular momentum discrepancy can not be explained by the retrograde envelope. Therefore, while we
have to be careful when interpreting our surface angular velocities, this
likely doesn't affect these particular results.  We should note that while our time
  averaged $\Omega_c/\Omega_e$ are similar to observed values it is
  worth keeping in mind that observations represent an instant in
  time, not a time average, so any of the values seen in
  Fig.\ref{histogram} could potentially be observed. Similarly, while
  we have been able to reproduce the rotation profile of KIC10526294
  with a time averaged profile, we expect that run long enough, our simulated
  profile would evolve.  

\section{Discussion}
Based on these 2D numerical simulations we conclude that IGW can
explain current observations of core-envelope
differential rotation in main sequence stars with a convective core.  These results lead to a few
conclusions and predictions.  Low flux models with low rotation can
show a variety of differential rotation profiles, ranging from $\sim$1--5.
Low flux models with high rotation have rotation profiles closer to
uniform, though notably, not exactly.  High flux models with low rotation
generally show faster and counter-rotating envelopes.
Finally, high flux models of high rotation (which have not yet been
observed) will have envelopes spinning faster than their core,
but with positive sign.  In our simulations the transition between
fast and slow rotators (or retrograde versus prograde envelopes)
occurs $\sim 10^{-5}$ rad/s, but this will likely depend on the age
and mass of the star.  

Given that simply changing the convective flux by a factor of two in
our simulations can lead to significantly different
rotation profiles it is worth discussing what could lead to a
different convective flux in a real star, or more appropriately, a different
surface wave flux.  Of course, the first is
mass, with higher mass stars having higher luminosities and therefore,
higher convective fluxes.  The second is age, as a star evolves its
luminosity increases somewhat which could lead to enhanced convective
fluxes.  However, as the star ages it also develops a severe gradient in the
Brunt-Vaisala frequency at the convective-radiative interface due to
the chemical composition gradient left behind by converting Hydrogen
to Helium.  Such a gradient could act as a filter to waves
  propagating outward to the surface and thus reduce the surface wave
  flux, possibly causing a massive star to appear more like a low flux
  model.  It is hard to know how these two effects combined affect
the surface wave flux.  In reality, changes in both mass and age are
also accompanied by changes in the
stratification throughout the radiative region which could affect the effective propagation and
dissipation of waves, and hence the surface wave flux.  Any of these effects
could contribute to individual stars being better described by
different model parameters than initially expected and all of these
effects should be considered in future models and as more observations
become available.

This work, in addition to \cite{ar15}, are some of the first to make
direct comparisons between numerical hydrodynamic simulations and
observations. Such comparisons clearly require some caveats.
First and foremost, these simulations are carried out in
  2D.  We expect
  that IGW transport would be more efficient in 2D as waves are not
  able to spread out over the sphere and because 2D turbulence has an
  inverse cascade.  Therefore, we expect the timescales of angular
  momentum transport in these simulations to be shorter than in the
  actual star, but we can not say by how much.  This difficulty in
  extrapolating timescales is because mean flow development depends on
  velocity correlations and it is difficult to say how much more
  efficient these correlations are in 2D versus 3D. Furthermore, we do not
  know how wave transport will proceed at
  higher latitudes but we expect it to be less efficient than at the
  equator.  Therefore, the observations, which represent a latitudinal
  average, are likely a lower limit of our simulated equatorial differential
  rotation.  Finally, treating mass and evolutionary state as simply a
  change in flux is inadequate. 
    
We have shown that our numerical simulations of IGW can explain the
observed differential rotation profiles observed in intermediate and massive main
sequence stars.  Given the shortcomings of these simulations (2D,
increased viscosity, increased thermal diffusivity) it is surprising
that the results agree as well as they do.  This agreement is likely
due to the limited observational constraints and to the fact IGW
transport can vary significantly.  As observational
constraints become more numerous, more sophisticated simulations which
properly consider mass and age of individual stars and proper dimensionality will be necessary.  In
turn, we expect that additional observational constraints can be used
to constrain simulation parameters.  One robust conclusion from both the observations and the numerical
simulations is that stellar rotation is complex and can admit a
variety of profiles.

\bibliographystyle{apj}

\begin{thebibliography}{}


\bibitem[Aerts et al.(2003)]{aerts03} Aerts, C., Thoul, A., 
Daszy{\'n}ska, J., et al.\ 2003, Science, 300, 1926 

\bibitem[Aerts (2008)]{aerts08} Aerts,C., 2008, Proceedings of IAU,
  250, 237
\bibitem[Aerts \& Rogers (2015)]{ar15} Aerts,C. \& Rogers, T.~M. 2015,
  ApJL, 806, L33
\bibitem[Ausseloos et al.(2004)]{aus04} Ausseloos, M., Scuflaire, R.,
  Thoul, A. and Aerts, C., 2004, MNRAS, 355, 352
 
\bibitem[Baldwin et al.(2001)]{baldwin01} Baldwin, M.~P., Gray, L.,
  Dunkerton, T.~J., Hamilton, K., Haynes, P.~H., Randel, W.~J.,
  Holton, J.~R. et al., 2001, Rev. Geophys., 39, 179

\bibitem[Beck et al.(2012)]{beck12} Beck, P.~G., Montalban, J.,
  Kallinger, T. et al., 2012, Nature, 481,55

\bibitem[Briquet et al.(2007)]{briquet07} Briquet, M., Morel, T.,
  Thoul, A., Scuflaire, R., Miglio, A., Montalban, J., Dupret, M.~A,
  Aerts, C., 2007, MNRAS, 381, 1482

\bibitem[Brown et al.(2012)]{brown12} Brown, B.~P., Vasil, G.~M. and
  Zweibel, E.~G., 2012, ApJ, 756, 109

\bibitem[Deheuvels et al.(2015)]{deheuvels15} Deheuvels, S., Ballot,
  J., Beck, P.~G., Mosser, B., Ostensen, R., Garcia, R.~A. and Goupil,
  M.~J., 2015, A\&A, 580, 96

\bibitem[Dupret et al.(2004)]{dupret04} Dupret, M.~A., Thoul, A.,
  Scuflaire, R., Daszynska-Daszkiewicz, Aerts, C., Bourge, P.~O.,
  Waelkens, C. and Noels, A., 2004, A\&A, 415,251

\bibitem[Eggleton (1971)]{egg71} Eggleton, P.~P., 1971, MNRAS, 151, 351

\bibitem[Fuller et al.(2014)]{fuller14} Fuller, J., Lecoanet, D.,
  Cantiello, M. and Brown, B., 2014, ApJ, 796, 17

\bibitem[Gough (1969)]{gough69} Gough, D.~O., 1969, J. Atmos. Sci.,
  26, 448

\bibitem[Heger \& Langer (2000)]{heger00} Heger, A. and Langer, N.,
  2000, ApJ, 544, 1016

\bibitem[Kurtz et al.(2015)]{kurtz15}  
Kurtz D.~W., Shibahashi H., Murphy S.~J., Bedding T.~R., Bowman D.~M., 
2015, MNRAS, in press (arXiv:1504.04245)

\bibitem[Mosser et al.(2012)]{mosser12} Mosser, J., Goupil, M.~J.,
  Belkacem, K. et al., 2012, A\&A, 548, 10

\bibitem[Pamyatnykh et al.(2004)]{pam04} Pamyatnykh, A.~A., Handler,
  G. and Dziembowski, W.~A., 2004, MNRAS, 350, 1022

\bibitem[P{\'a}pics et al.(2014)]{Papics2014}
  P{\'a}pics P.~I., Moravveji E., Aerts C., Tkachenko A., Triana S.~A., Bloemen
  S., Southworth J., 2014, A\&A, 570, A8

\bibitem[P{\'a}pics et al.(2015)]{Papics2015}
  P{\'a}pics P.~I., Tkachenko A., Aerts C., Van Reeth T., De Smedt K., Hillen
  M., \O stensen R., Moravveji E., 2015, ApJ, 803, L25

\bibitem[Saio et al.(2015)]{saio15} Saio, H., Kurtz, D.W., Takata, M., Shibahashi, H., Murphy, S.~J.,
Sekii, T. and Bedding, T.~R., 2015, MNRAS, 447, 3264

\bibitem[Rogers \& Glatzmaier
    (2005)]{rg05} Rogers T.~M. and Glatzmaier, G.~A., 2005, ApJ, 620, 432

\bibitem[Rogers et al.(2013)]{rg13}
    Rogers T.~M., Lin D.~N.~C., McElwaine J.~N., Lau H.~H.~B., 2013,
    ApJ, 772, 26
\bibitem[Schmid et al.(2015)]{valentina15} Schmid, V.~S., Tkachenko,
  A., Aerts, C., Degroote, P., Bloemen, S. et al. 2015, A\&A, in press

\bibitem[Talon et al.(1997)]{talon97} Talon,
  S., Zahn, J.~P., Maeder, A. and Meynet, G., A\&A, 322,209

\bibitem[Triana et al.(2015)]{triana15}
Triana S.~A., Moravveji, E., Papics, P.~I., Aerts, C., Kawaler,
S.~D. and Christensen-Dalsgaard, J., 2015, ApJ, 810, 16

\bibitem[Zahn (1992)]{zahn92} Zahn, J.~P.,
  1992, A\&A, 265,115

\end{thebibliography}

\acknowledgments 
Support for this research was provided by NASA grant NNX13AG80G to
T.Rogers.  Computing was carried out on Pleiades at NASA
Ames. T. Rogers would like to thank an anonymous referee for
insightful and helpful comments which improved this manuscript
significantly.  She would also like to thank C.Aerts, S. Triana and E. Morraveji for useful
conversations leading to the development of this manuscript.


\end{document}